\documentclass[prb,a4paper,10pt,twocolumn,floatfix,superscriptaddress,amsfonts,latin1,utf8]{revtex4-2}
\pdfoutput=1
\usepackage[T1]{fontenc}\usepackage[latin1]{inputenc}
\usepackage{dcolumn,graphicx,color,booktabs,microtype,afterpage} 
\usepackage{sidecap}
\usepackage{times}
\usepackage[colorlinks,plainpages=false,linkcolor=blue,urlcolor=blue,citecolor=blue,pdfpagemode=UseNone,pdfstartview=FitBH]{hyperref}

\usepackage{amssymb}
\usepackage{amsmath}
\usepackage{graphicx}% Include figure files
\usepackage{esvect}
\usepackage{bm}% bold math
\usepackage{tabularx}
\usepackage{stackengine}
\stackMath

\usepackage{graphicx}% Include figure files
\usepackage{dcolumn}% Align table columns on decimal point
\usepackage{bm}% bold math
\usepackage{siunitx}
\usepackage{chemformula}
\usepackage{float}
\let\ce\ch

\newcommand{\CNO}{\ce{CoNb2O6}}
\newcommand{\COP}{\ce{Co^{2+}}}

\begin{document}

\preprint{APS/123-QED}

\title{Single-ion properties of the transverse-field Ising model material CoNb$_2$O$_6$}
\pacs{xx.xx.mm}

% Don't forget to put in \email{} for each
\author{J. A. Ringler}
\affiliation{%
Department of Physics, Colorado State University, Fort Collins, CO 80523, USA
}
\author{A. I. Kolesnikov}
   \affiliation{Neutron Scattering Division, Oak Ridge National Laboratory, Oak Ridge, Tennessee 37831, USA}
\author{K. A. Ross}
    \affiliation{%
Department of Physics, Colorado State University, Fort Collins, CO 80523, USA
}%
    \affiliation{Quantum Materials Program, CIFAR, MaRS Centre,
West Tower 661 University Ave., Suite 505, Toronto, ON, M5G 1M1, Canada}

\date{\today}% It is always \today, today,

\begin{abstract}
\CNO\ is one of the few materials that is known to approximate the one-dimensional transverse-field Ising model (1D-TFIM) near its quantum critical point.  It has been inferred that Co$^{2+}$ acts as a pseudo-spin 1/2 with anisotropic exchange interactions that are largely Ising-like, enabling the realization of the TFIM.  However, the behavior of \CNO\ is known to diverge from the ideal TFIM under transverse magnetic fields that are far from the quantum critical point, requiring the consideration of additional anisotropic, bond-dependent (Kitaev-like) terms in the microscopic pseudo-spin 1/2 Hamiltonian.   These terms are expected to be controlled in part by single-ion physics, namely the wavefunction for the pseudo-spin 1/2 angular momentum doublet.   Here, we present the results of both inelastic neutron scattering measurements and electron paramagnetic resonance spectroscopy on \CNO, which elucidate the single-ion physics of Co$^{2+}$ in \CNO\ for the first time.  We find that the system is well-described by an intermediate spin-orbit coupled Hamiltonian, and the ground state is a well-isolated Kramers doublet with an anisotropic $g$-tensor.  We provide the approximate wavefunctions for this doublet, which we expect will be useful in theoretical investigations of the anisotropic exchange interactions.

\end{abstract}

\maketitle

%%% INTRODUCTION SECTION
\section{\label{sec:Intro}Introduction}
The transverse-field Ising model (TFIM) is arguably the simplest and most tractable model that exhibits a quantum phase transition, yet it leads to a wealth of intriguing results \cite{2015_Dutta_Book}. Of particular current interest, this model has been used to predict non-equilibrium phenomena such as Kibble-Zurek scaling, quantum annealing in glassy spin systems, and potential violations of the eigenstate thermalization hypothesis and the emergence of the subsequent many-body localized phase \cite{2020_Oshiyama_JPSJ,2019_Silevitch_NatCommun,2014_Kjall_PhysRevLett}.  Materials that realize the TFIM are highly sought-after for both testing of such theoretical ideas and materials engineering. However, the requirement of having a low enough (e.g. laboratory-accessible) transverse magnetic field strength to induce the quantum critical point (QCP) strongly limits the number of TFIM materials.  There are only three known solid-state magnetic systems which approximate the TFIM, including dipolar-coupled 3D \ce{LiHoYF4} \cite{2011_Gingras_JPhysConfSer}, as well as quasi-1D systems \ce{(Ba/Sr)Co2V2O8} \cite{2019_Cui_PhysRevLett} and {\CNO} {\cite{2010_Coldea_Science}}, the latter being the system of interest in this study. 

\begin{figure}[t]
    \includegraphics{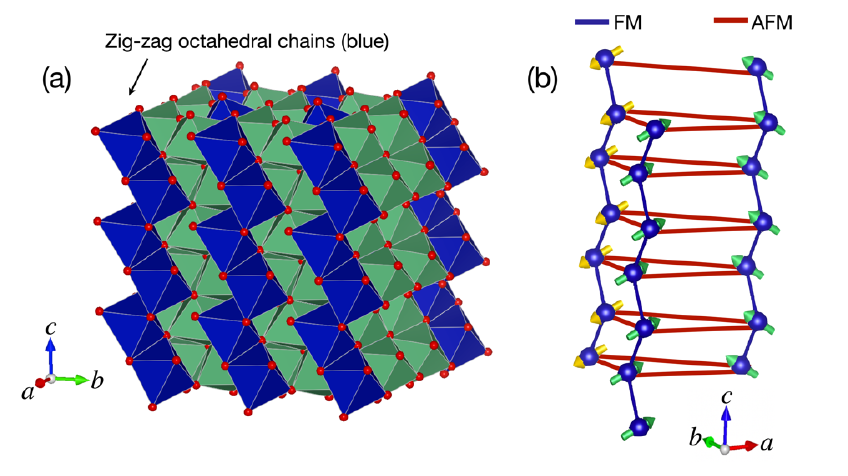}
    \caption{Visualization of the crystal and magnetic structures of {\CNO}. (a) For {\CNO} in the columbite structure, the {\COP} is octahedrally coordinated with \ce{O^{2-}} ions, which run in a zig-zag pattern along the \textit{c}-axis, as described in the main text. (b) The effective spin-$\frac{1}{2}$ moments orient in the \textit{ac}-plane and form an isosceles triangular network with AFM exchange in the \textit{ab}-plane.}
    \label{fig:crys_mag}
\end{figure}

The quasi-1D TFIM magnet {\CNO} is known to exhibit signatures of a 1D QCP \footnote{It should be noted that the 3D transition actually occurs at $\mu_0 H_c=\SI{5.5}{T}$, as observed by inelastic neutron scattering measurements \cite{2010_Coldea_Science}. However, the 1D QCP at zero temperature can be inferred from properties of the quantum critical fan \cite{2014_Kinross_PhysRevX}.} under an applied transverse field strength of $
\mu_0 H_c = \SI{5.25}{T}$, as observed via various thermodynamic measurements such as specific heat and magnetization, as well as nuclear magnetic resonance \cite{2014_Kobayashi_PhysRevB,2016_Kobayashi_PhysRevB,2015_Liang_NatCommun, 2014_Kinross_PhysRevX}. {\CNO} belongs to a family of $3d$ transition metal niobates \ce{$M$Nb2O6} which form in the columbite structure (space group P\textit{bcn} (60), $M$ = \ce{Cd}, \ce{Co}, \ce{Cu}, \ce{Fe}, \ce{Mn}, \ce{Ni}, \ce{Zn}) \cite{2017_Munsie_Thesis, 2009_Pullar_JAmCeramSoc}, where zig-zag chains of edge-sharing \ce{$M$}-\ce{O^{2-}} octahedra run along the crystallographic \textit{c}-axis (Fig.\ref{fig:crys_mag}(a)). For {\COP} in this structure, the crystal electric field (CEF) of the local environment and spin-orbit coupling (SOC) conspire to form effective spin-$\frac{1}{2}$ degrees of freedom with magnetic moments that tend to lie in the \textit{ac}-plane (in the absence of a transverse field).  These pseudo-spins  form strongly-coupled ferromagnetic (FM) chains along \textit{c} \cite{1977_Maartense_SolidStateCommunications}, the details of which are discussed in Sec. \ref{sec:Modeling}. The chains are linked together antiferromagnetically (AFM) in the \textit{ab}-plane (Fig. \ref{fig:crys_mag}(b)), and the \textit{b}-axis is perpendicular to the net  ordered moments.  The pseudospins have been assumed to be coupled via Ising interactions, a model that correctly reproduces many of the features near the critical point with a field along the transverse axis.  
 
However, away from the critical point there have been some notable deviations from ideal TFIM behavior, which until recently had been explained only phenomenologically \cite{1995_Heid_JournalofMagnetismandMagneticMaterials,2014_Kjall_PhysRevLett,2000_Kobayashi_PhysRevB,2010_Coldea_Science,2014_Robinson_PhysRevB,Fava2020}.  Recently there has been a renewed interest in the microscopic details of the Hamiltonian in the regime far from the QCP. A recent study by Fava \textit{et al.} \cite{Fava2020} demonstrated that the symmetry being broken near the QCP is not that of a global Ising symmetry, but that of a \textit{glide} symmetry which is a consequence of its 3D space group. This symmetry analysis showed that there are off-diagonal terms in the Hamiltonian beyond the usual Ising exchange that are needed to account for all discrepancies between the model and experimental data.
Shortly thereafter, Morris \textit{et al.} \cite{2021_Morris_NatPhys} modeled their time-domain terahertz (THz) spectroscopy data with a Hamiltonian that explained the overall ordered moment direction as the average of two different observed Ising exchange axes in the $ac$ plane, which alternate along the \textit{c}-axis.  These bond-dependent Ising interactions inspired the authors to form a new microscopic Hamiltonian the authors dubbed the ``twisted'' Kitaev chain, which is another model of fundamental importance in non-equilibrium physics and contains off diagonal interactions equivalent to those proposed for \CNO in Ref. \onlinecite{Fava2020}.  Additional THz measurements supported their findings and ultimately led to the possible first direct measurement of Kramers-Wannier duality in nature \cite{1941_Kramers_PhysRev}. In order to match their field-dependent terahertz data, Morris \textit{et al.} proposed an orientation of the Ising exchange axes; the best fit had the same ordered moment direction in the $ac$-plane, but additionally included a component that alternated $\pm$ 17 degrees along the $b$-axis.  However, their fit did not take into account the $g$-tensor, which should play an important role in this determination, since it combines with the Ising exchange axes to determine the final ordered moment direction.   

In both of these recent works proposing terms beyond the simple Ising ferromagnet, the models map back to the TFIM and produce quantum critical behavior within the same Ising universality class as the 1D TFIM.  However, questions remain about these new Kitaev-like terms that become increasingly relevant farther from the QCP.

Preceding these efforts in \CNO, there has been a growing interest in understanding bond-dependent anisotropic interactions (such as Kitaev interactions) in $3d$ transition metal compounds, particularly {\COP}-based magnetic systems \cite{Liu2020KitaevCompounds,2020_Vivanco_PhysRevB,2020_Zhong_ScienceAdv}. Liu \textit{et al.} recently showed that $3d$ cobaltates should display bond-dependent interactions, and in some situations the Kitaev term would dominate the microscopic Hamiltonian \cite{Liu2020KitaevCompounds}. These calculations were applied to ``honeycomb lattices'', formed by layers of edge-sharing, trigonally-distorted \ce{CoO6} octahedra. The authors parameterized  the {\COP} CEF ground state wavefunctions by the trigonal distortion strength, which determines the strength of the non-Kitaev terms and leads to a rich phase diagram. Although \CNO\ is not a honeycomb lattice material, and the octahedra are not simply trigonally distorted, the work of Liu \textit{et al.} inspires the notion that the details of the Kitaev-like interactions in {\CNO} can be understood by studying the single-ion wavefunctions of {\COP} in the \CNO\ lattice.   This is part of the motivation for this work.

In this article, we present the results of inelastic neutron scattering (INS) measurements on pure {\CNO} which used a wide incident neutron energy range $E=60-\SI{2500}{meV}$, and use an intermediate spin-orbit coupled model to determine its single-ion energy levels (Sec. \ref{sec:single_ion_results}) as well as the wavefunctions of the {\COP} CEF ground state doublet (App. \ref{app:cef_gr_st}).  We then present the EPR spectrum measured from a diluted variant (described in Sec. \ref{sec:Methods}). These data conspire to definitively show that the {\ce{Co^{2+}}} single-ion ground state is a well-isolated Kramers doublet with strongly anisotropic moments described by a rhombic $g$-tensor. Although this scenario is expected based on prior work, this study provides the details that were previously missing, and which can be used to develop a theoretical understanding of the anisotropic exchange in {\CNO}; namely a direct determination of the transverse field $g$-value,  $g_b \approx 3.3$, and the details of ground state CEF wavefunctions.

The contents of the paper are structured as follows: In Sec. \ref{sec:Methods}, we describe the experimental details of all measurements. In Sec. \ref{sec:Modeling}, we explain the single-ion models used to fit the INS and EPR data. In Sec. \ref{sec:Results}, we present the results and discuss them in the context of the single-ion model, from which the conclusions are drawn in Sec. \ref{sec:Conclusions}.

%%%%% METHODS
\section{\label{sec:Methods}Experimental Methods}

In order to investigate the large dynamic range expected for the \ce{Co^{2+}} single-ion energy levels, we used the SEQUOIA time-of-flight chopper spectrometer in the Spallation Neutron Source at Oak Ridge National Laboratory for the INS experiment \cite{2010_Granroth_sequoia}.  This instrument is optimized for magnetic neutron scattering studies, with ample low-$|\mathbf{Q}|$ detector coverage, and allows a wide range of incident neutron energies.  The experiment was performed on an 8.62 g powder sample of pure \ce{CoNb2O6}, which we synthesized via a standard solid state synthesis as reported elsewhere \cite{2011_Sarvezuk_JournalofAppliedPhysics}.  The powder was loaded into a cylindrical aluminum sample can under helium exchange gas. Data were taken with a wide range of incident energies ($E_i = 60, 150, 250, 700$ and \SI{2500}{meV}) at two temperatures where appropriate ($T =\SI{5}{K}$ and $T = \SI{200}{K}$). Instrumental settings and configurations, such as the $T_0$ and Fermi chopper (FC1 and FC2) and elastic energy resolutions for each of the incident neutron energies, can be found in the appendix. 

We acquired the low-temperature ($T = \SI{5}{K}$) EPR spectra using a Bruker EMX Elexsys spectrometer equipped with a ColdEdge closed-cycle helium cryostat. We used a signal with  microwave frequency $f=\SI{9.6453}{GHz}$ and a modulating field amplitude of $A_{mod}=\SI{0.4}{mT}$, corresponding to a microwave power of $P=\SI{6.325}{mW}$. For EPR measurements, we substituted 1\% Co into the isomorphic non-magnetic analog \ce{MgNb2O6} system (\ce{Mg_{0.99}Co_{0.01}Nb2O6}). This created magnetically isolated Co$^{2+}$ ions in the the same crystal field environment as in \ce{CoNb2O6}, ensuring that spin-spin interactions did not interfere with the determination of the single-ion $g$-values. The spectra were analyzed using the EasySpin toolbox for MATLAB \cite{EasySpin}, where we simulated a spectrum that was calculated by defining a system consisting of a pseudospin $(S_{\text{eff}}=1/2)$ (described in Sec. \ref{sec:EPR}) and a single \ce{^{59}Co} nucleus, which has a natural abundance of 1.
%%%% SINGLE-ION MODEL 
%%%% NEED TO FIX SECTION TITLE
\section{\label{sec:Modeling} Single-Ion Modeling}

In this section we present the methods used in modeling the single-ion data throughout this study, and were used to fit the data presented in Sec. \ref{sec:Results}. We first introduce the intermediate spin-orbit coupled Hamiltonian used in calculating the \ce{Co^{2+}} single-ion energy levels, as well as the resulting INS response. This method, described in Sec. \ref{sec:SIH}, has been demonstrably effective in $3d$ transition metal compounds, particularly in octahedrally-coordinated \ce{Co^{2+}} complexes \cite{2017_Ross_PhysRevB,2019_Sarte_PhysRevB,2011_Tomiyasu_PhysRevB,1971_Holden_JPhysC} and is analogous to those typically used in rare-earth magnets \cite{2012_Abragam_Book, Jensen_Book_RareEarthMagnetism}; the difference being in that for the rare-earth case the single-ion Hamiltonian is formulated firmly in the $J=||L+S||$ basis, rather than treating the spin-orbit coupling term on equal-footing with the CEF term. We then describe the EPR Hamiltonian. The resulting phenomenology is that the single-ion \ce{Co^{2+}} levels in a distorted octahedron are Kramers doublets with magnetic dipole moments described by an anisotropic g-tensor. 
\begin{figure}
    \includegraphics{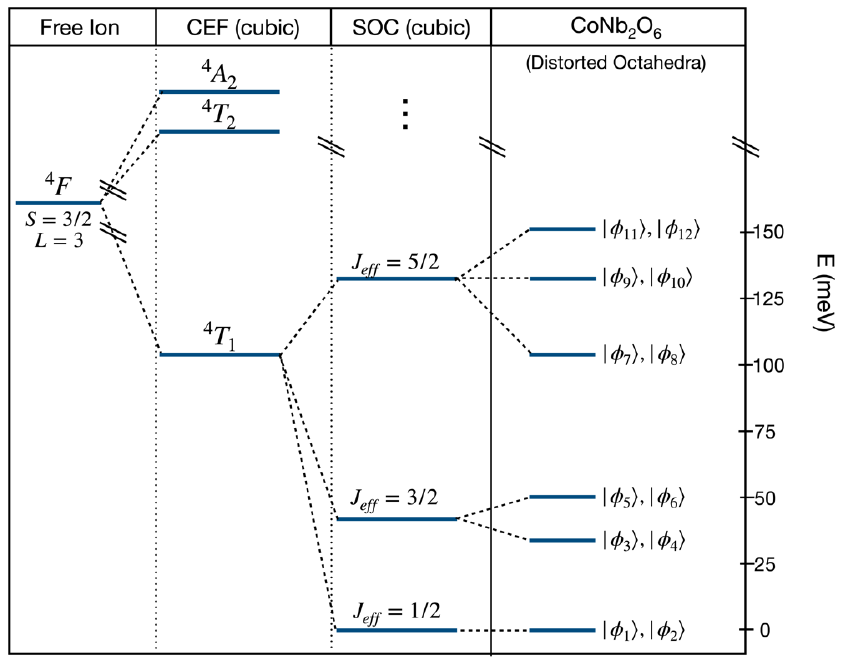}
    \caption{Energy level scheme depicting the splitting due to perturbations on the $^4F$ free-ion ground state. Note that the left three panels in this diagram corresponds to a purely cubic CEF (i.e., an octahedral field with no distortions). The cubic CEF splits the free-ion ground state into three orbital multiplets; the ground-state $^4T_1$ triplet is well-separated in energy from the excited states \cite{1957_Kanamori_ProgTheorPhys}. The ground state degeneracy is reduced further into three spin-orbit-split multiplets, which is further split into 14 doublets. The energy eigenvalues of the lowest doublets resultant from the INS fits are displayed in the right panel (see Table \ref{tab:single_ion_levels} for a complete and specific list of all energy eigenvalues).}
    \label{fig:energy_splitting}
\end{figure}

\subsection{\label{sec:SIH} Single-ion Hamiltonian}
We begin by noting that for a $3d^7$ transition metal ion Hund's rules dictate that the \ce{Co^{2+}} free ion ground state is $^4F$ $(S=3/2, L=3)$, leading to a $28\times 28$ Hilbert space. The odd number of unpaired electrons ($d^7$) implies all eigenstates are at least doubly degenerate, per Kramers degeneracy theorem. As is common for $3d$ transition metals {\cite{2012_Abragam_Book}}, we approximate the single-ion Hamiltonian ($\hat{\mathcal{H}}_{\text{SI}}$) to consist of a contribution from the crystal electric field and a contribution from spin-orbit coupling (SOC) as follows,

\begin{equation}
    \hat{\mathcal{H}}_{\text{SI}}(\mathbf{L},\mathbf{S}) = \hat{\mathcal{H}}_{\text{CEF}}(\mathbf{L}) + \hat{\mathcal{H}}_{\text{SOC}}(\mathbf{L},\mathbf{S})
    \label{eqn:single_ion_H}
\end{equation}

\noindent where the CEF Hamiltonian $\hat{\mathcal{H}}_{\text{CEF}}(\mathbf{L})$ can be written in Stevens form \cite{2012_Abragam_Book}, 

\begin{equation}
    \hat{\mathcal{H}}_{\text{CEF}}(\mathbf{L}) = \sum_{l,m}B_l^m\hat{O}_l^m 
\end{equation}

\noindent The $B_l^m \, (\hat{O}_l^m)$ are the Stevens CEF parameters (operators). For transition metal ions, the Stevens operators are written in terms of the orbital angular momentum operators $L_z,L_+,L_-$ {\cite{1964_Hutchings_SolStatePhys}}. Point charge calculations reveal that only eight CEF parameters are nonzero, due to the $C_2$ point group symmetry of the \ce{Co^{2+}} average local environments. CEF levels are inherently magnetic, and thus can be probed using INS.  In Sec. \ref{sec:single_ion_results}, we use INS data to determine the Steven's parameters.

In the intermediate (Russell-Saunders) spin-orbit coupling scheme, the spin-orbit interaction is given by

\begin{equation}
    \hat{\mathcal{H}}_{\text{SOC}}(\mathbf{L},\mathbf{S}) = \lambda \mathbf{L} \cdot \mathbf{S} = \lambda \sum_\alpha \hat{L}_\alpha \hat{S}_\alpha, 
\end{equation}

\noindent where $\lambda = \SI{-22.32}{meV}$ for the free \ce{Co^{2+}} ion {\cite{1957_Kanamori_ProgTheorPhys}} and for $\alpha = (x,y)$, the orbital angular momentum operators $\hat{L}_\alpha$ are expressed in terms of the usual raising and lowering operators: $\hat{L}_x = \frac{1}{2} (\hat{L}_+ + \hat{L}_-)$ and $\hat{L}_y = \frac{1}{2i} (\hat{L}_+ - \hat{L}_-)$, and likewise for $\hat{S}_x$ and $\hat{S}_y$.

%as well as an excited orbital singlet $(^4A_2)$ and triplet $(^4T_2)$ for the perfect octahedron; a cartoon energy level diagram depicts the splittings in Fig. \ref{fig:energy_splitting}. Treating the SOC Hamiltonian as a perturbation on the CEF and diagonalizing again causes the ground state orbital triplet to split into a ground state doublet $(J_{\text{eff}} = 1/2)$, and two excited levels in the form of a quartet and sextet $(J_{\text{eff}} = 3/2$ and $J_{\text{eff}} = 5/2)$. 
Diagonalizing the CEF Hamiltonian keeping only terms consistent with cubic symmetry (i.e. a perfect octahedron) results in a ground state orbital triplet $(^4T_1)$ that is well-separated from the excited multiplets \cite{2019_Sarte_PhysRevB}, as seen in Fig. \ref{fig:energy_splitting}. From there, SOC and crystalline distortions further lower the degeneracy resulting in a ground state doublet $(J_\text{eff}=\frac{1}{2})$. In {\CNO}, the ground state doublet is well isolated in energy (by $E\approx\SI{30}{meV} \approx \SI{350}{K}$), as observed in the INS data presented in Sec. \ref{sec:single_ion_results}, allowing for the description of this system as an \textit{effective} spin-1/2 system at sufficiently low temperatures. The components of the $g$-tensor can then be extracted in zero field as the matrix elements of the magnetic dipole moment operator in the subspace of the ground state wave functions $|\phi_n\rangle$ ($n=1,2$):
\begin{equation*}
\label{eqn:gtens}
    g_{i} = 2 \langle \phi_n | \hat{L}_i + 2 \hat{S}_i | \phi_n \rangle ,
\end{equation*}
Strictly speaking, the ``$g$-tensor'' is not actually a tensor (it does not transform properly under rotations), and these $g$-values are those obtained from the square root of the eigenvalues of the $g$-tensor and correspondingly, the principal axes are the eigenvectors of $G$ (see Ref. \cite{2012_Abragam_Book}). 

\subsection{\label{sec:INS} Inelastic neutron scattering response}
The single-ion dynamic structure factor (DSF) associated with Eqn. (\ref{eqn:single_ion_H}) at constant momentum transfer $|\mathbf{Q}|$ is calculated in the following form \cite{Jensen_Book_RareEarthMagnetism}:
\begin{equation}
    S(E) = C\sum_{n,m,\alpha} \frac{e^{-\beta E_n}}{Z} \frac{\Gamma_{(n,m)} |\langle \phi_n | \hat{L}_\alpha + 2\hat{S}_\alpha | \phi_m \rangle |^2}{(E_n - E_m - E)^2 + \Gamma_{(n,m)}^2},
    \label{eqn:DSF}
\end{equation}
where $\beta=1/k_BT$, $Z$ is the partition function, $|\phi_n\rangle$ is the wave function associated with energy eigenvalue $E_n$ from the single-ion calculations, and $\Gamma_{(n,m)}$ is a Lorentzian half-width at half-maximum (HWHM) that can account for spectral broadening due to dispersion, finite lifetimes in excited states, inhomogeneous broadening, and approximate instrumental resolution. The actual observed intensity is related to the DSF through $I(|\mathbf{Q}|,E)=\frac{k_f}{k_i}f^2(|\mathbf{Q}|)S(|\mathbf{Q}|,E)$ \cite{1996_Squires_Book}, where $k_f,k_i$ are the final and incident wave numbers of neutrons that scatter with corresponding energy transfer $\Delta E = \frac{\hbar^2}{2m}(k_i^2 - k_f^2)$ and $f(|\mathbf{Q}|)$ is the magnetic form factor. At constant $|\mathbf{Q}|$, the form factor is also constant and is assimilated into the constant $C$ in Eqn. (\ref{eqn:DSF}), which acts as an overall scale factor. 

\subsection{\label{sec:EPR} EPR Hamiltonian}
The spin Hamiltonian used in EPR spectral analysis is approximated by two contributions due to the electron Zeeman interaction and the hyperfine coupling between the $3d^7$ electrons and the \ce{^{59}Co} nuclear spin ($I=7/2$). It is important to emphasize that here these interactions are written in terms of  the effective spin $S_{\text{eff}}$ rather than the ``bare'' spin. For systems with strong single-ion anisotropy, the interaction terms are written in the following tensor form \cite{2018_Piwowarska_AppMagnRes},
\begin{equation}
    \hat{\mathcal{H}}_{\text{spin}}(\mathbf{S}_{\text{eff}},\mathbf{I} \,) = -\mu_B {\mathbf{S}_{\text{eff}}\cdot\hat{g}\cdot {\mathbf{B}} + \hbar {\mathbf{S}_{\text{eff}}} \cdot \hat{A} \cdot {\mathbf{I}}}
    \label{eqn:epr_hamiltonian}
\end{equation}
\noindent where $\mathbf{B}$ is the magnetic field responsible for driving the EPR transition, and $\hat{g}, \, \hat{A}$ are the $g$ and hyperfine coupling tensors, respectively. The eigenvalues of these tensors are readily extracted from fitting to EPR spectral data. For this work, the absorption lineshape is assumed to be Lorentzian. 

%%%%% RESULTS
\section{\label{sec:Results}Results and Discussion}
%In this section, we present the results of both INS and EPR measurements. We first discuss the INS results, from which the {\COP} single-ion energy levels were calculated and discuss the results in the context of the single-ion model used to fit the data. From there, the EPR spectrum is presented, from which the $g$-values are directly determined and compared to results from other measurements. 

\subsection{\label{sec:single_ion_results}Single-ion energy levels from INS}
\begin{figure*}[!th]
\includegraphics{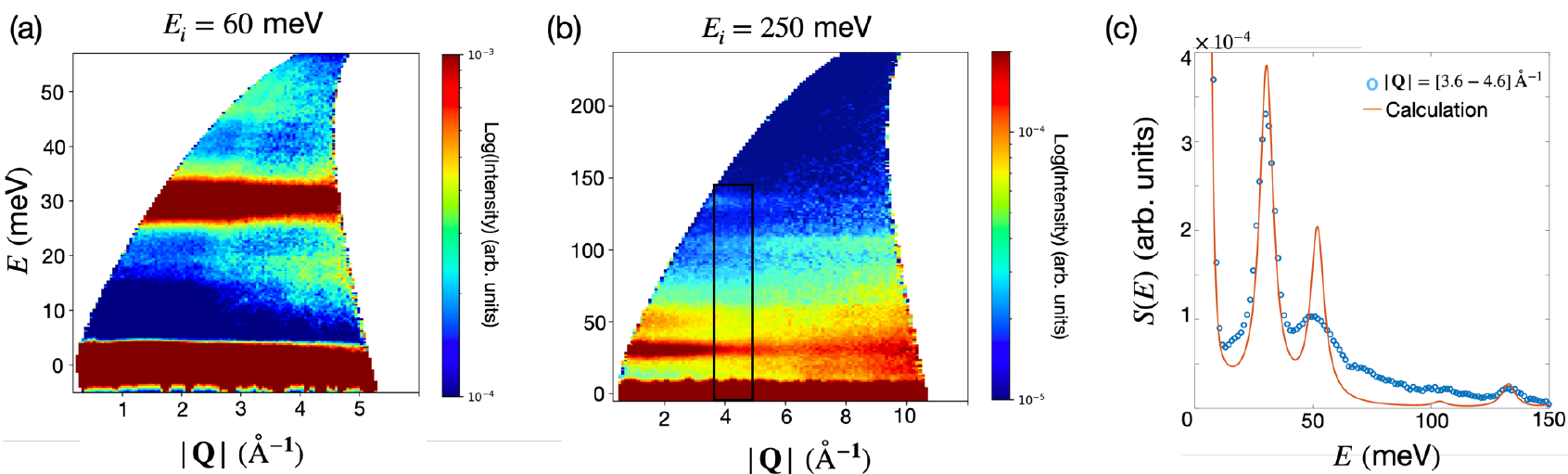}
\caption{\label{fig:INS_slicecutfit} Dynamic structure factor $S(|\mathbf{Q}|,E)$ data, measured via INS on a \CNO\ powder sample. (a) Data at $T=5$ K using $E_i = 60$ meV neutrons. The absence of any low-lying CEF levels below $E=\SI{30}{meV}$ indicates the ground state is a well-isolated doublet at temperatures well below those at which the first excited state is thermally populated $(T\ll \SI{350}{K})$. (b) Data at $T=5$ K using  $E_i = 250$ meV neutrons, displaying the transition energies of the first four excited doublets.  Prominent phonon contributions are seen at high $|\mathbf{Q}|$, spanning from the elastic line to just above $E=100$ meV. (c) Constant-$|\mathbf{Q}|$ cut through the data in panel (b), along with the calculated fit based on the single-ion model. Note that the fit includes a peak at $E=105$ meV that is not visible in the data; the calculated intensity is relatively small, so the peak is likely disguised by the phonon background.}
\end{figure*}

The single-ion levels of {\COP} were observed as peaks corresponding to transitions in the single-ion DSF acquired from INS experiments on the SEQUOIA time-of-flight spectrometer. These were fit to a CEF Hamiltonian, and the resulting calculated energy levels ranged from 0 to $\SI{902}{meV}$.  Only a few of these modes were observed via INS due to weak CEF level intensities and large background at the higher energy transfers; a complete list of all calculated and observed energies is found in Table \ref{tab:single_ion_levels}, along with the calculated intensities (normalized to the $\SI{30}{meV}$ peak) for each of the modes. An asterisk denotes the mode was not observed, while a dagger indicates that there were some weak magnetic signatures in constant-$E$ cuts (see Fig. \ref{fig:app_104_peak}). The modes and their intensities that \textit{were} observed, along with the $g$-values from EPR, provide sufficient information to obtain the Steven's parameters from fits.  The DSF at $T=$5 K for incident neutron energies $E_i=60, \, \SI{250}{meV}$ is presented in Fig. \ref{fig:INS_slicecutfit}, along with an intensity versus energy transfer cut and fit for $E_i=\SI{250}{meV}$.  All data presented in this article are background subtracted to remove the contributions from the sample environment, including the aluminum sample can.  These ``empty-can'' datasets were collected at all specified incident neutron energies and sample environment temperatures. The remaining background results from sample-specific features, such as coherent and incoherent phonon scattering.   All data presented here were corrected by a factor of $k_f/k_i$ which means that the quantity being displayed in Fig. \ref{fig:INS_slicecutfit} is proportional to $f^2(|\mathbf{Q}|)S(|\mathbf{Q}|,E)$ (in arbitrary units).

\begin{table}[ht]
\caption{\label{tab:single_ion_levels}
Calculated single-ion {\COP} energy levels in {\CNO} from the fit to the INS data shown in Fig. \ref{fig:INS_slicecutfit} compared to observed approximate energy levels. Note that not all of the transitions listed were observed in the data, as is further discussed in the main text. Calculated and observed intensities (relative to the $\SI{30}{meV}$ peak) are displayed in the far right column, and are separated by horizontal lines that indicate incident neutron energies used in the calculation ($E_i=250,700,\SI{2500}{meV}$, from top to bottom). Unobserved modes have an asterisk but a dagger indicates there were weak magnetic signatures in constant-$E$ cuts (Fig. \ref{fig:app_104_peak} in App. \ref{app:mag_modes}).} 
\begin{ruledtabular}
\begin{tabular}{ccccc}
\textrm{Eigenstate}& $E_m$ & Obs. & Calc. Int. & Obs. Int. \\
\colrule
$\phi_1,\phi_2$ & $0$ & 0 & - & -\\
$\phi_3,\phi_4$ & $30.4$ & $29.8(8)$ & 1 & 1  \\
$\phi_5,\phi_6$ & $51.9$ & $49(2)$ & 0.30 & 0.0181\\
$\phi_7,\phi_8$ & $104$ & $*\dagger$ & 0.024 & *\\
$\phi_9,\phi_{10}$ & $133$ & $133(1)$ & 0.076 & 0.0039 \\
$\phi_{11},\phi_{12}$ & $150$ & * & 0.0007 & *\\
\colrule
$\phi_{13},\phi_{14}$ & $377$ & $*$ & 0.060 & *\\
$\phi_{15},\phi_{16}$ & $390$ & $*$ & 0.056 & *\\
$\phi_{17},\phi_{18}$ & $467$ & $*$ & 0.052 & *\\
$\phi_{19},\phi_{20}$ & $477$ & $*$ & 0.052 & *\\
$\phi_{21},\phi_{22}$ & $493$ & $*$ & 0.041 & *\\
$\phi_{23},\phi_{24}$ & $511$ & $*$ & 0.027 & *\\
\colrule
$\phi_{25},\phi_{26}$ & $898$ & $*\dagger$ & 0.040 & *\\
$\phi_{27},\phi_{28}$ & $902$ & $*\dagger$ & 0.039 & *\\
\end{tabular}
\end{ruledtabular}
\end{table}

Our data analysis consisted of fitting the single-ion model described in Sec. \ref{sec:INS} to the $E_i = \SI{250}{meV}$ data, which showed the clearest features. A Gaussian peak at the elastic line with full-width at half-maximum (FWHM) taken from the instrument resolution function was also added to account for both coherent and incoherent nuclear elastic scattering. Then, the instrument resolution function was further used to provided the FWHM in fitting the inelastic peaks to a Lorentzian profile. Overall, nine parameters were fit, including the eight CEF parameters $(B_l^m)$ and the overall scale factor $C$.  The data contained 7 clear pieces of information; 3 peak positions and 4 peak intensities (the intensity of the peak near 100 meV is an upper limit since the position of said peak is difficult to discern); thus the fit would be underconstrained. To remedy this, we added a penalty for the distance of the calculated principal $g$-values from the INS model, to those measured via EPR (reported in Sec. \ref{sec:epr_results}). Fits were accomplished by first employing a particle swarm optimization method using the point charge calculation as a staring point, then once the algorithm found a ``global'' minimum, we used nonlinear least-squares to explore the region around the minimum and estimate the uncertainties of the parameters. As seen in Fig. \ref{fig:INS_slicecutfit}(c), there is a broad signal underlying the CEF peaks that we did decided not to attempt to model phenomenologically, since it still represents constraint on peak positions and intensities.  This ``background'' may have contributed to the optimizer's tendency to get stuck in local minima without identifying the strong peaks in the data.  To improve the tendency to find the peaks, we added an additional penalty for the distance between the nearest calculated transitions and the observed peak positions. This was accomplished by first fitting the peaks individually to extract a peak position, and then incorporating a least-squares term to match the calculated peak centers to the observed (this is partially accomplished by fitting the full intensity profile, but this gives it an extra boost for matching peak positions). A similar procedure used in an example fit program that is provided by the software package PyCrystalField \cite{PyCrystalField_2018}. The final results of the fitted parameters (except for the scale factor) are presented in Table \ref{tab:ins_parameters}.
%\REF[appendix]
The first excited level was detected using an incident neutron energy of $E_i = \SI{60}{meV}$ was at $E=\SI{30}{meV}$, and is typical for octahedrally-coordinated {\COP} compounds \cite{2017_Ross_PhysRevB, 2011_Tomiyasu_PhysRevB, 1971_Holden_JPhysC}.  In such systems it is attributable to the transition between $J_{\text{eff}}=1/2$ and $3/2$ multiplets \cite{2011_Tomiyasu_PhysRevB,1971_Buyers_JPhysC}. The level was observed at both temperatures $(T=5, \, \SI{200}{K})$, and displayed a decrease in intensity as momentum transfer $|\mathbf{Q}|$ increases, which is characteristic of magnetic excitations.   

\begin{table}[!ht]
\caption{\label{tab:ins_parameters}%
Stevens parameters extracted from fitting to the inelastic neutron scattering data (the {\COP} single-ion DSF, Eqn. ({\ref{eqn:DSF}})).  } 
\begin{ruledtabular}
\begin{tabular}{ccccc}
&\textrm{Parameter}& &
\textrm{Value}$(\SI{}{meV})$&\\
\colrule
&$B_{2,0}$ & & $-1.14(6)$&\\
&$B_{2,1}$ & & $0.1(2)$&\\
&$B_{2,2}$ & & $6(2)$&\\
&$B_{4,0}$ & & $-0.78(4)$&\\
&$B_{4,1}$ & & $-0.55(5)$&\\
&$B_{4,2}$ & & $0.15(5)$&\\
&$B_{4,3}$ & & $0.330(1)$&\\
&$B_{4,4}$ & & $3.29(4)$&\\

\end{tabular}
\end{ruledtabular}
\end{table}

Some spectral features calculated from the fit deviated from the observed modes in various ways. For example, the mode at $E=\SI{51.9}{meV}$ is observed to be wider and much less intense than calculated but is appropriately positioned. We attribute this to possible structural disorder, as has been detailed by Sarkis \textit{et al.} in recent AC susceptibility work on {\CNO} \cite{2021_Sarkis_PhysRevB}. Such inhomogeneities would lead to a range of CEF levels and would broaden observed CEF transitions. 

The $\SI{51.9}{meV}$ peak corresponds to transitions only between the ground state and excited states, as the first excited level is not appreciably thermally populated, since it lies at energy $E=\SI{348}{K}/k_B$ above the ground state and orbital selection rules forbid intra-multiplet transitions. This however does not exclude the presence of mixing between CEF and exchange interactions, as the dominant FM exchange energy ($J_0\approx \SI{3}{meV}$ \cite{Fava2020}) is only about an order of magnitude smaller than the CEF splittings and can possibly lead to dispersion that can widen the signal. In this case (which is seen in both \COP systems studied in Ref. \cite{2017_Ross_PhysRevB}), these dispersive effects would be invisible to us since the energies are smaller than the SEQUOIA instrument resolution for the incident neutron energies used (found in App. \ref{app:exp_settings}).

There is then the absence of a clear peak near $E=\SI{104}{meV}$ which is suggested by the single-ion calculation. Looking with a loving eye there is perhaps a slight peak visible there, and signatures of a magnetic excitation were seen in constant-$E$ cuts of the $E_i=\SI{150}{meV}, T=5,\SI{200}{K}$ data; the same is true for the highest energy modes (see App. \ref{app:mag_modes}). While the masking of lower-energy excitations could potentially be simply explained as phonon background, the higher-energy cases are not so simple. In this regime the modes could possibly be concealed by more complex background signals such as those found due to multi-phonon processes and recoil scattering. This observed recoil scattering (see Fig. \ref{fig:app_104_peak} in Appendix \ref{app:mag_modes}) is plausibly attributed to the presence of hydrogen, because sharp peaks were found in the $E_i=\SI{2500}{meV}$ data at energies that are consistent with the various \ce{O-H} bending and stretching modes displayed in water molecules and hydroxyl groups and suggests that some water had accumulated on the sample's surface.

\subsection{\label{sec:cef_gs_gtensor}Calculated CEF ground state and $g$-tensor}
The spin and orbital composition of the CEF ground state doublet $|\phi_{1,2}\rangle$ obtained from the analysis of the INS data is listed in Table \ref{tab:CEF_GS} located in Appendix \ref{app:cef_gr_st}. The largest contributions come from the $L=0,1,3$ subspaces.   The approximate wavefunctions for the ground state doublet in the $|L_z, S_z \rangle$ basis are:
\begin{equation}
\label{eqn:cef_gs}
    |\phi_{1,2}\rangle \approx \mp 0.63 |\pm 3, \pm \frac{3}{2}\rangle \pm 0.53 |0, \pm \frac{1}{2}\rangle \pm 0.38 |\mp 1, \pm \frac{3}{2}\rangle \pm ... 
\end{equation}

The unabbreviated form of Eqn.(\ref{eqn:cef_gs}) is what is used to calculate the $g$-tensor of the system from the INS results, the principal values of which are located in Table \ref{tab:epr_parameters}. After INS data were fit, the obtained Stevens parameters were adjusted within their respective 95\% confidence intervals to find how much the resulting $g$-values vary, the range of which is roughly equivalent to the 95\% CI (2$\sigma$) of the $g$-values. From the ground state wavefunction we can also find the principal axes of the $g$-tensor relative to the crystallographic directions, and find that the strongest value ($g_z=6.4$) is along an axis in the \textit{ac}-plane, approximately $37^{\circ}$ from \textit{c} along \textit{a}. 
\begin{table}[ht]
\caption{\label{tab:epr_parameters}%
Parameters from Eqn. (\ref{eqn:epr_hamiltonian}) extracted by fitting to EPR signal acquired from the dilute sample and include the principal $g$-values ($g_{\alpha}$, unitless and located above the line), the hyperfine coupling values $(A_{\alpha})$, and the Lorentzian linewidth $\gamma$. For the $g$-values, the values obtained via EPR are displayed next to those calculated from the fit to the INS data (and their reasonable range) as described in Sec. \ref{sec:SIG}. The furthest right column shows the percent difference between the two methods.}
\begin{ruledtabular}
\begin{tabular}{cccc}
\textrm{Parameter}&
\textrm{EPR}& INS & \% Difference \\
\colrule
$g_{x}$ & 3.33 & 3.9(6) & 16\% \\
$g_{y}$ & 3.01 & 3.3(8) & 9.2\% \\
$g_{z}$ & 6.39 & 6.3(6) & 1.4\% \\
\hline
$A_x$ & \SI{123}{\mu eV} & &  \\
$A_y$ & \SI{71.5}{\mu eV} & &  \\
$A_z$ & \SI{521}{\mu eV} & &  \\
$\gamma$ & \SI{5.3}{mT} & &  
\end{tabular}
\end{ruledtabular}
\end{table}
This is reasonably close to the observed ordered moment direction of $31^{\circ}$ in the \textit{ac}-plane \cite{1979_Scharf_JournalofMagnetismandMagneticMaterials,2016_Kobayashi_PhysRevB}. We note that the observed ordered moment direction is not necessarily required to be the same as the $g$-tensor principal axes; ultimately, it is a combination of the exchange anisotropy and the $g$-tensor anisotropy that dictates the moment direction; one clear example of this is the splayed ferromagnetism found in systems like \ce{Yb2Ti2O7} and \ce{Yb2Sn2O7} \cite{2016_Gaudet_PhysRevB,2013_Yaouanc_PhysRevLett}.  As for the final results of the fitted Stevens parameters, it was found that the confidence interval for $B_2^1$ straddled the value of zero, and setting the parameter equal to zero had negligible effect on the $g$-tensor and the calculated intensity pattern. In fact, when $B_2^1$ is entirely removed from the fitting routine it results in the same energy spectrum, but slightly changes the $g$-values (all three differ from the reported results by less than 6\%) and rotates the principal axes by $3^\circ$ in the \textit{ac}-plane, as well as reducing the error on the parameters by at least an order of magnitude. This could all indicate that $B_2^1$ is an extraneous parameter that is not necessary to describe this system. We stress that this does not necessarily imply a higher point group symmetry; the next highest symmetry ($C_{2v}$) would require additional parameters to vanish.

\subsection{\label{sec:epr_results}EPR spectral analysis}
\begin{figure}
    \includegraphics{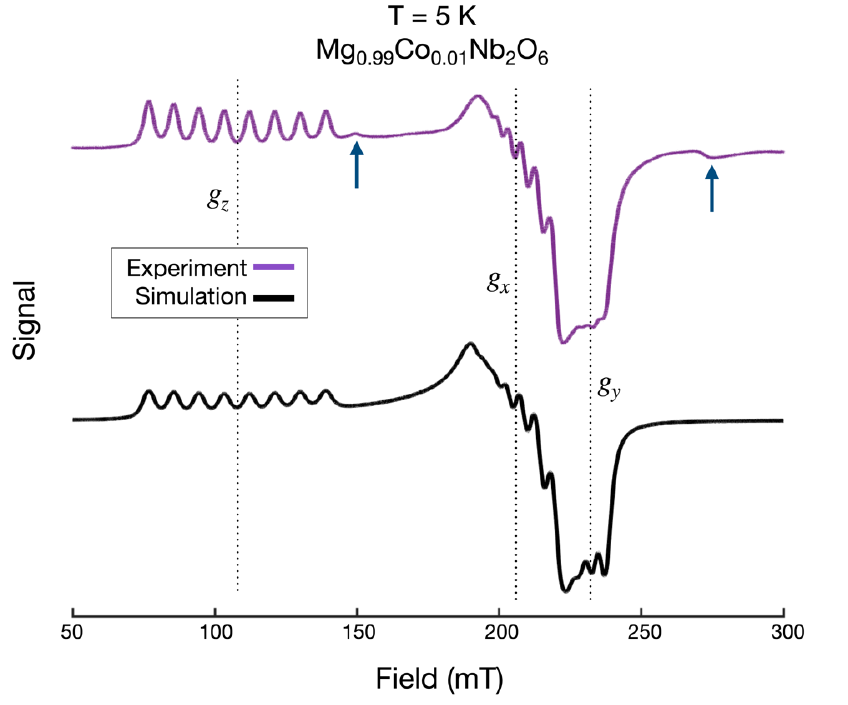}
    \caption{Field-swept X-band EPR response at $T=5$ K, compared to a simulated system. The g-values can be estimated from the center positions of the major features. The spacing between the jagged peaks correspond to hyperfine splittings. The cause of the spectral features at approximately $B=\SI{150,275}{mT}$ (indicated by blue arrows) remain unknown.}
    \label{fig:EPR_Signal}
\end{figure}

The low-temperature X-band EPR spectrum is displayed in Fig. \ref{fig:EPR_Signal}.  This measurement was done on a powder sample, so the directions of the principal axes of the $\hat{g}$ and $\hat{A}$ could not be directly measured, but the principal values $(g_{ii}, A_{ii})$ extracted from the fit are displayed in Table \ref{tab:epr_parameters}. The existence of 3 separate principal $g$-values is consistent with a rhombic $g$-tensor. This is in contrast to an axial system $(g_z=g_\parallel, \, g_x=g_y=g_\perp)$ which is usually associated with Ising-like single-ion systems. Two of the $g$-values are close but not equal, leading to an overlap of spectral features associated with the values. There is a noticeable difference in the clarity of the hyperfine-split features between the experiment and simulated signals, which is likely due to insufficient instrument resolution. Microwave power and modulation amplitudes were adjusted in an effort to resolve the splittings. Note the two features at $B=150$ and \SI{275}{mT} in the experimental data which are not accounted for in the simulation and the source(s) remain unkown.

The principal $g$-values from the EPR spectrum follow the same general trend as those determined from the CEF calculations ($g_z>g_y>g_x$), and the values for $g_z$ (associated with the Ising moment direction) differ by approximately $0.6\%$. The reason for the larger differences between the values of $g_x$ and $g_y$ is likely compositional, i.e., the INS measurements were performed on pure \CNO, and EPR spectra were acquired from the extremely dilute sample. However, we find it important to make a certain point more salient: the "most correct" $g$-values are likely the ones found via EPR, as the dilute sample provided an environment that would be much closer to true single-ion behavior.  However, it is also true that the sample is primarily MgNb$_2$O$_6$, which will have slightly different structural parameters than CoNb$_2$O$_6$,  so these g-values are still an approximation to the ``true'' single-ion behavior of CoNb$_2$O$_6$.

As mentioned previously, the EPR spectrum was approximated using a pseudospin-1/2 model, which means the zero-field splitting (ZFS) terms usually present in $S>1/2$ systems (like \ce{Co^{2+}} when the full $S=3/2$ $L=3$ manifold is included \cite{2018_Piwowarska_AppMagnRes}) were omitted from Eqn. (\ref{eqn:epr_hamiltonian}). Additionally, the \ce{^{59}Co} nucleus has a large quadrupolar moment $(Q/|e|=0.42\times 10^{-24} \, \si{cm^{-2}}$), so a quadrupole interaction term should also be included. However, the absence of both ZFS and quadrupole terms had negligible effect on the simulation.  This is consistent with another octahedrally-coordinated \ce{Co^{2+}} compound, discussed in the Supplementary Information of Ref. \cite{2014_Gomez-Coca_NatCommun}. Hyperfine interaction strengths are $\sim \SI{}{\mu eV}$ and therefore cannot be corroborated with our INS data due to having insufficient energy resolution. A Lorentzian FWHM of $\gamma = \SI{5.3}{mT}$ accounted for the linewidth due to inhomogeneous broadening from unresolved hyperfine features. 

%%%%%CONCLUSIONS
\section{\label{sec:Conclusions}Conclusions}
We have presented inelastic neutron scattering and electron paramagnetic resonance results on the Ising material {\CNO}. The main results are the measurements and analysis of the {\COP} single-ion properties of this system; primarily the single-ion energy levels, the ground state CEF wavefunction, and the $g$-values. The ground state of {\COP} in these octahedrally-coordinated environments is a Kramers doublet which is well-separated from the first excited states by over $\SI{300}{K}$. These results show that the system is well-described by an intermediate spin-orbit coupled Hamiltonian. Further calculations are needed to explore the Kitaev-like nature of the interactions in {\CNO}, which can be accomplished via similar methods to those used in Ref.\cite{Liu2020KitaevCompounds} and using the CEF ground state wavefunctions and principal $g$-values that we have provided. 
%%%%% ACKNOWLEDGMENTS
\begin{acknowledgments}
This research was supported by the U.S. Department of Energy, grant no. DE-SC0018972. The authors would like to gratefully acknowledge the discussions with Joe Zadrozny and Richard Miller. The neutron scattering data were reduced using the Mantid software package \cite{2014_Arnold_NIM}, while data analysis utilized PyCrystalField and MATLAB \cite{MATLAB:2010,PyCrystalField_2018}. The authors also acknowledge the Molecular and Materials Analysis Center of the Analytical Resources Core at Colorado State University for instrument access and training. A portion of this research at Oak Ridge National Laboratory's Spallation Neutron Source was sponsored by the Scientific User Facilities Division, Office of Basic Energy Sciences, U.S. Department of Energy. 
\end{acknowledgments}

\appendix

\section{\label{app:cef_gr_st} CEF ground state wavefunctions and g-tensor orientation}
\begin{figure}[h!]
    \includegraphics{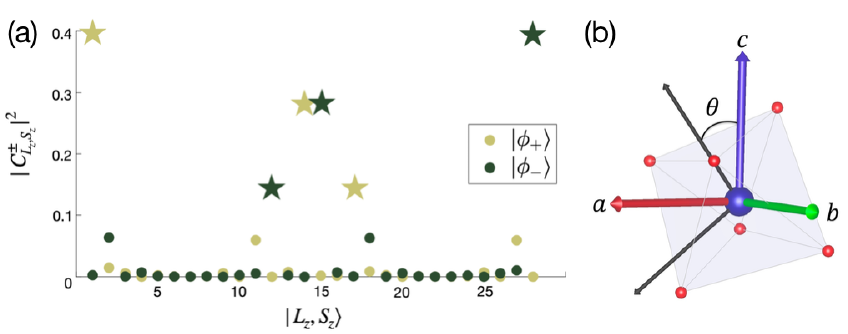}
    \caption{(a) A plot of the absolute square amplitudes of the CEF ground state doublet. The horizontal axis is made up of integers that correspond to the indices of contributions from the eigenstates of the $|L_z,S_z\rangle$ basis as found in Table \ref{tab:CEF_GS}. The wavefunctions are highly symmetric. Stars indicate the greatest contributions. (b) Depiction of the $g$-tensor principal axes (black vectors) as calculated by the methods described in Sec. \ref{sec:SIH} referenced to the crystallographic directions, where two axes are rotated approximately $\theta=37^{\circ}$ from the \textit{c}-axis in the \textit{ac}-plane. The third principal axis is constrained by symmetry to lie along \textit{b}. }
    \label{fig:CEF_spectrum}
\end{figure}

The wavefunctions for {\COP} are expressed as linear combinations of eigenstates in the intermediate coupling $(\lvert L_z,S_z \rangle)$ basis as follows:

\begin{equation}
    \label{eqn:CEF_WF}
    \lvert \phi_n \rangle = \sum_{L_z,S_z} C^{(n)}_{L_z,S_z} \lvert \, L_z,S_z \rangle
\end{equation}
Where $L_z,S_z$ are the eigenvalues of the orbital and spin angular momentum operators $\hat{L}_z$ and $\hat{S}_z$, respectively. The $C^{(n)}_{l,s}$ are the amplitudes of the corresponding eigenstates, which can be found in Table \ref{tab:CEF_GS}. A plot of the absolute square amplitudes is shown in Fig. \ref{fig:CEF_spectrum}, and demonstrates the highly symmetric nature of the ground state. Knowing the composition of the ground state wavefunctions in the pseudospin-1/2 regime is one key to understanding the possible Kitaev nature of the microscopics in coordinated {\COP} complexes. 

\begin{table}[!ht]
\caption{\label{tab:CEF_GS}List of amplitudes $C^{(n)}$ corresponding to the single-ion {\COP} ground state doublet wavefunctions $(n=1,2)$ in the $\lvert L_z,S_z \rangle$ basis. See Eqn. \ref{eqn:CEF_WF}. Lefthand column is the index of the contributions from the eigenstates of the $|L_z,S_z\rangle$ basis, and is important for understanding Fig. \ref{fig:CEF_spectrum}.}
\begin{ruledtabular}
\begin{tabular}{ccccc}
\vspace{1mm} Index & $L_z$ & $S_z$ & $C_{L_z,S_z}^{(1)}$ & $C_{L_z,S_z}^{(2)}$\\
\colrule 
1 & -3 & -3/2 & 0.0498 & 0.6290 \\  
2 & -3 & -1/2 & -0.2516 & -0.1213 \\
3 & -3 & 1/2 & 0.0185 & 0.0765 \\   
4 & -3 & 3/2 & -0.0841 & -0.0048 \\ 
5 & -2 & -3/2 & -0.0324 & -0.0483 \\
6 & -2 & -1/2 & 0.0165 & 0.0234 \\  
7 & -2 & 1/2 & -0.0161 & -0.0168 \\ 
8 & -2 & 3/2 & 0.0278 & -0.0255 \\  
9 & -1 & -3/2 & -0.0012 & -0.0764 \\
10 & -1 & -1/2 & 0.0497 & -0.0093 \\ 
11 & -1 & 1/2 & 0.0731 & -0.2433 \\  
12 & -1 & 3/2 & 0.3791 & 0.0044 \\   
13 & 0 & -3/2 & 0.0458 & 0.0878 \\   
14 & 0 & -1/2 & -0.0016 & -0.5293 \\ 
15 & 0 & 1/2 & 0.5308 & 0.0404 \\    
16 & 0 & 3/2 & 0.0844 & -0.0390 \\   
17 & 1 & -3/2 & -0.0257 & -0.3799 \\ 
18 & 1 & -1/2 & 0.2499 & 0.0927 \\   
19 & 1 & 1/2 & -0.0132 & -0.0506 \\  
20 & 1 & 3/2 & 0.0766 & 0.0048 \\    
21 & 2 & -3/2 & 0.0278 & 0.0299 \\   
22 & 2 & -1/2 & -0.0156 & 0.0148 \\  
23 & 2 & 1/2 & -0.0221 & 0.0147 \\   
24 & 2 & 3/2 & -0.0459 & 0.0286 \\   
25 & 3 & -3/2 & 0.0019 & 0.0840 \\   
26 & 3 & -1/2 & -0.0752 & 0.0125 \\  
27 & 3 & 1/2 & -0.1017 & 0.2428 \\   
28 & 3 & 3/2 & -0.6270 & -0.0000 \\  
\end{tabular}
\end{ruledtabular}
\end{table}

Furthermore, it is important to explain the coordinate system regarding the $g$-tensor principal axes as calculated from the single-ion Hamiltonian (Eqn. \ref{eqn:single_ion_H}). As stated in the main text, the fitting routine used starting parameters from a point charge calculation in which the coordinate axes were aligned along the crystallographic axes, as in, ($x\parallel a, y\parallel b, z\parallel c$). This means that the final fit parameters are also expressed in that coordinate system, where the $g$-tensor is not diagonal but rotated an angle $\theta=37^{\circ}$ about the \textit{b}-axis, which is consistent with the observed moment direction but differs by a few degrees (See Fig. \ref{fig:CEF_spectrum} below). We wish to reiterate that the observed moment direction is a combination of both $g$-tensor and exchange anisotropy. 

\section{Signatures of unobserved magnetic modes and recoil scattering}
\label{app:mag_modes}
\begin{figure}[h!]
    \includegraphics{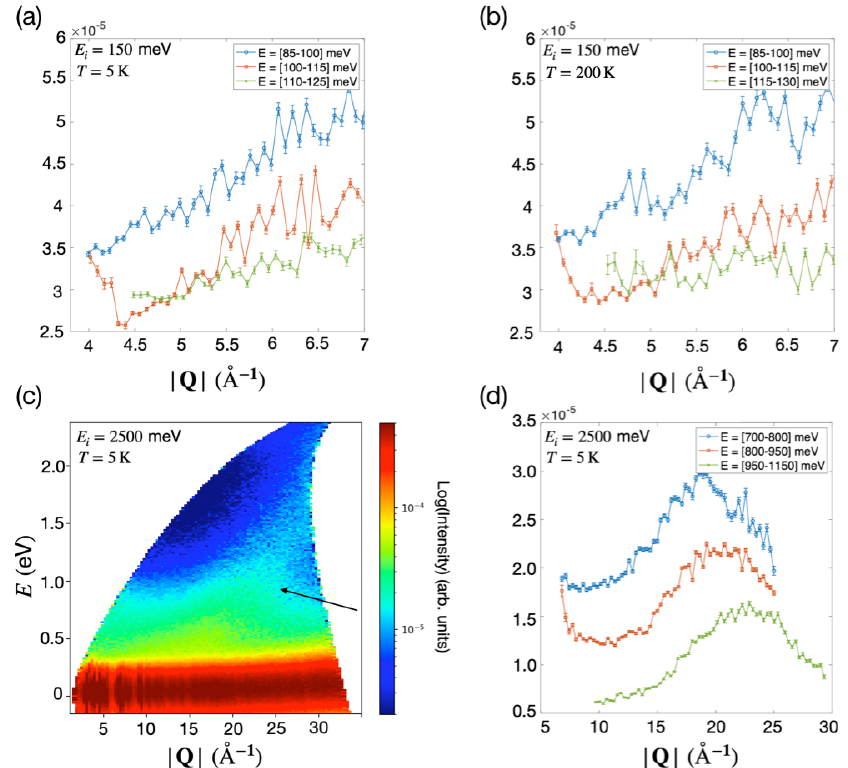}
    \caption{Magnetic origins of several modes unobserved in constant-$|\bf{Q}|$ cuts. (a) Constant-$E$ cuts of the data with incident neutron energy $E_i=\SI{150}{meV}$ at $T=\SI{5}{K}$, demonstrating the magnetic nature of the $\SI{104}{meV}$ mode. (b) Similar cuts for the same incident neutron energy but at $T=\SI{200}{K}$, corroborating the results in (a). (c) Slice of the $E_i=\SI{2500}{meV}$ data taken at $T=\SI{5}{K}$. The arrow indicates where strong recoil scattering is observed (as mentioned in Sec. \ref{sec:single_ion_results}), and likely masks much of the intermediate modes. (d) Cuts of the data in panel (c) showing the magnetic signatures of the excitations near $\SI{900}{meV}.$}
    \label{fig:app_104_peak}
\end{figure}

As mentioned in the main text, not all modes were observed. Modes unobserved in constant-$|\bf{Q}|$ cuts such as the one shown in Fig. \ref{fig:INS_slicecutfit} can be identified as being magnetic by observing their $|\bf{Q}|$-dependence at constant energy transfer; the magnetic form factor $f(|\bf{Q}|)$ causes the intensity to drop off with increasing momentum transfer. This is seen in Fig. \ref{fig:app_104_peak}(a,b,d), which displays constant-$E$ cuts of the $E_i=150,\SI{2500}{meV}$ data to expose the magnetic origins of the $104,898$, and $\SI{901}{meV}$ modes. 

\section{\label{app:exp_settings}Instrumental configurations used on SEQUOIA}

A list of configurational settings for the SEQUOIA spectrometer used in this research is listed in Table {\ref{tab:sequoia_settings}}. 

\begin{table}[!hbt]
\caption{\label{tab:sequoia_settings}%
Instrumental configuration settings for the SEQOUIA spectrometer for all incident neutron energies $(E_i)$ and sample environment temperatures. Includes the frequency settings for the $T_0$ and high flux/resolution Fermi choppers (FC1/FC2), as well as the elastic energy resolution $\delta E$.} 
\begin{ruledtabular}
\begin{tabular}{ccccc}
$E_i (\SI{}{meV})$ & $T (\SI{}{K})$ & $T_0 (\SI{}{Hz})$ & FC $(\SI{}{Hz})$ & $\delta E (\SI{}{meV})$ \\
\colrule
60 & 5, 200 & 90 & FC2, 420 & 1.61 \\
150 & 5, 200 & 90 & FC2, 600 & 3.98 \\
250 & 5, 200 & 120 & FC2, 600 & 6.95 \\
700 & 5 & 150 & FC1, 600 & 50.73 \\
2500 & 5 & 180 & FC1, 600 & 334.75
\end{tabular}
\end{ruledtabular}
\end{table}

\providecommand{\noopsort}[1]{}\providecommand{\singleletter}[1]{#1}%

\end{document}